# Research and experimental verification on low-frequency long-range underwater sound propagation dispersion characteristics under dual-channel sound speed profiles in the Chukchi Plateau

Jinbao Weng[1,a], Yubo Qi[2], Yanming Yang[1], Hongtao Wen[1], Hongtao Zhou[1], Ruichao Xue[1]

1. Laboratory of Ocean acoustics and Remote Sensing, Institute of Oceanography, Ministry of Natural Resources, Xiamen, Fujian 361005, China
2. State key laboratory of acoustics, Institute of Acoustics, Chinese Academy of Sciences, Beijing 100190, China

## ABSTRACT

The dual-channel sound speed profiles of the Chukchi Plateau and the Canadian Basin have become current research hotspots due to their excellent low-frequency sound signal propagation ability. Previous research has mainly focused on using sound propagation theory to explain the changes in sound signal energy. This article is mainly based on the theory of normal modes to study the fine structure of low-frequency wide-band sound propagation dispersion under dual-channel sound speed profiles. In this paper, the problem of the intersection of normal mode dispersion curves caused by the dual-channel sound speed profile (SSP) has been explained, the blocking effect of seabed terrain changes on dispersion structures has been analyzed, and the normal modes has been separated by using modified warping operator. The above research results have been verified through a long-range seismic exploration experiment at the Chukchi Plateau. At the same time, based on the acoustic signal characteristics in this environment, two methods for estimating the distance of sound sources have been proposed, and the experiment data at sea has also verified these two methods.

## I. INTRODUCTION

With global warming of the sea, in recent decades, the Chukchi Plateau has formed a unique dual-channel SSP with the joint formation of Pacific intrusion water and Atlantic water layers. If the sound source is located in the lower channel of the dual-channel SSP, the sound signal can reduced reflection with sea ice, thereby reducing energy attenuation and achieving long-range propagation. This excellent sound propagation characteristic is conducive to long-range sound source detection and positioning, as well as underwater acoustic communication, making sound propagation under dual-channel a research hotspot in Arctic Acoustics.

Currently, many experiments have been conducted to study sound propagation under dual-channel SSPs. A one-year sound propagation experiment was conducted in the Canadian Basin of the Arctic from 2016 to 2017, the experimental data showed that the spatiotemporal variation of the dual-channel SSP and the sea ice cover can cause the fluctuation of sound propagation loss to reach 60 dB. ICEX conducted low-frequency sound propagation experiments in the Canadian Basin in 2018. During this experiment, a horizontal array was used to receive low-frequency sound signals of different frequencies emitted by transducers. In the article, the corresponding relationship between the vertical distribution of normal mode eigenfunctions at different frequencies and the vertical distribution of dual-channel SSP was used to explain the trend of sound propagation loss with respect to the frequency of the sound source. The key point is that, when the eigenfunction is mainly distributed in the lower channel of the dual-channel SSP, the sound signal can achieve long-distance propagation.

[a] Email: wengjinbao@tio.org.cn

In many acoustic experiments conducted in the Arctic, sound receiving equipment often collects long-distance seismic exploration air gun sound signals. Keen et al. conducted research on the acoustic signals of air gun in the deep and shallow waters of the Arctic using the seismic exploration acoustic signals collected in the Beaufort Sea in August 2009 and the shallow waters of the Chukchi Sea from August to October 2013. Overall, the main focus of this work is to analyze the attenuation of acoustic signal intensity with propagation distance and the dispersion structure of acoustic signals in these two marine environments. However, this work did not further analyze the dispersion structure of acoustic signals, nor did it carry out work on normal mode separation and extraction of dispersion structures.

In the past decade, the application of warping transformation in underwater acoustics has gradually emerged. For the acoustic propagation conditions in the Arctic marine environment, the current main approach is to used the warping transformation of reflected normal modes to process acoustic signals such as whales. However, under the unique deep-sea SSP conditions in the Arctic, refracted normal modes are easily formed, and there is currently no work on performing warping transformation for modal separation of acoustic signals in the Arctic. For the mode separation of acoustic signals in the Arctic deep-sea environment, the main method used in the past was to use large aperture synchronous vertical arrays. However, the difficulty of deploying enlarged aperture vertical arrays in the deep sea is relatively high, especially under conditions such as the Arctic.

Yang conducted research on the dispersion structure of acoustic signals in the Arctic SSP environment in 1984 and proposed a pulse signal localization method based on the Arctic sound signal dispersion structure. Compared the forty years ago, the marine environment in the Arctic has undergone significant changes. One of these changes is the appearance of significant dual-channel SSP in the Canadian Basin and Chukchi Plateau, which may have new impacts on the dispersion structure of low-frequency sound signals. This article will analyze this. The remainder of this paper is organized as follows. Section II describes the basic theory of normal mode in ocean acoustics and the warping transformation of refractive normal modes. Section III describes the process of the sound propagation experiment at the Chukchi Plateau and the marine environment during the experiment. Section IV describes the simulation of sound propagation and received sound signals at the Chukchi Plateau, including eigenfunctions, time-domain waveforms, and dispersion structures. Section V describes the warping transformation and modal separation of the air gun acoustic signal, as well as the time-domain waveform and dispersion structure of the received acoustic signal with different propagation ranges and different reception depths. Section VI describes two methods for estimating sound source range based on the time-domain waveform of sound signals in a dual-channel SSP, as well as corresponding experimental data validation. Finally, Section VII summarizes the results of this study.

## II. THEORYS

### A. Normal mode theory

According to the book of << Computational Ocean Acoustics >>, the ocean sound field can be represented by normal modes, and the sound pressure at the receiving point can be represented by a series of superposition of normal modes,

$$P(\omega,r,z) = S(\omega) \frac{je^{-j\pi/4}}{\rho(z_s)\sqrt{8\pi r}} \sum_{m=1}^{M} \Psi_m(z_s)\Psi_m(z_r) \frac{e^{-\alpha_m(\omega)r}e^{jk_{rm}(\omega)r}}{\sqrt{k_{rm}(\omega)}}. \quad (1)$$

Among them, $\omega$ represents the angular frequency, $k_{rm}(\omega)$ represents the horizontal wave number, $\Psi_m(z)$ represents the eigenfunction, $z_s$ and $z_r$ represents the source depth and the receiver depth, $r$ represents the distance between source and receiver, $M$ represents the number of effective normal modes, $S(\omega)$ represents the spectral level of the sound source. Assuming the spectral level of a broadband pulse sound source is 1, the sound signal pressure can be abbreviated as follows,

$$P(\omega,r,z) = \sum_{m=1}^{M} A_m(\omega) e^{jk_{rm}(\omega)r}. \quad (2)$$

Among them, $A_m(\omega) = \dfrac{e^{j\pi/4}\Psi_m(z_s)\Psi_m(z_r)e^{-\alpha_m(\omega)r}}{\rho(z_s)\sqrt{8\pi r k_{rm}(\omega)}}$.

### B. Warping transform for refractive normal mode

Qi et al. derived the horizontal wavenumber of refractive normal modes in shallow water waveguides and derived the corresponding warping transformation operator. For the Arctic deep-sea marine environment, after detailed derivation, it was found that the expression for the horizontal wave number needs to be modified, but the warping transformation operator for refractive normal mode in Arctic is consistent, as follows,

$$h(t) = t_r - t^{-2}. \tag{3}$$

The expression for the inverse operator is as follows,

$$h^{-1}(t) = (t_r - t)^{-1/2}. \tag{4}$$

## III. INTRODUCTION TO THE EXPERIMENTAL PROCESS

During the period from August 20, 2021 to September 10, 2021, the US polar exploration experiment vessel "Sikuliaq" conducted a large-scale seismic exploration experiment based on air gun sound sources in the Chukchi Plateau waters. At the same time, an acoustic submersible buoy deployed in the Chukchi Plateau waters received most of the remote low-frequency sound signals from seismic exploration lines. This article will conduct in-depth research based on this randomly collected set of air gun sound propagation data. The position and relative distance of the seismic exploration experiment vessel and the acoustic submersible buoy are shown in Figure 1.

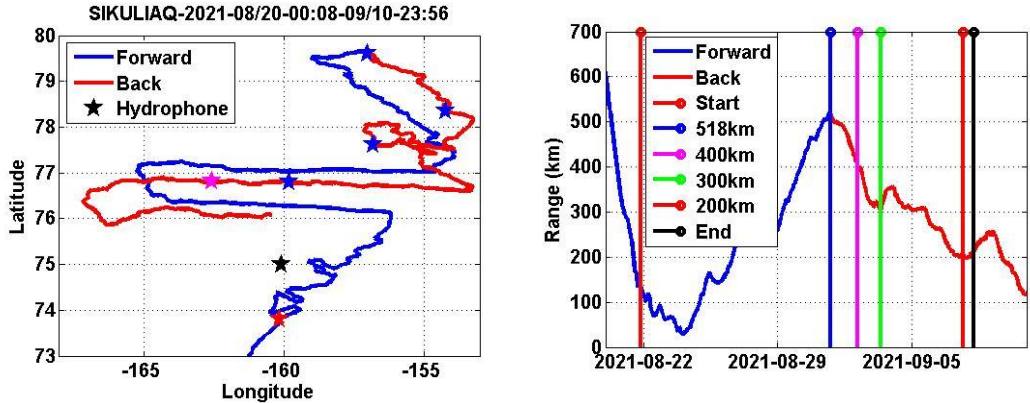

Figure 1 the trajectory of the seismic exploration experimental ship and the position of the acoustic submersible buoy, as well as the distance between them

### A. Signal receiving equipment

The acoustic submersible buoy system in this experiment consists of six self-contained underwater acoustic signal receiving devices, which are divided into three groups, namely two groups with spacing of 20 m. These three groups of devices are respectively distributed in the surface, middle, and bottom water bodies of the deep sea area of the Chukchi Plateau. The actual averaged depths of the specific devices are shown in Table 1.

Table 1 Expected and actual depths of acoustic receiving equipment

| Receiving equipment | Actual averaged depth |
|---|---|
| First group | 52m/67m |
| Second group | 883m/898m |
| Third group | 1662m/1677m |

## B. Seafloor topography

In this seismic survey experiment, according to the EOP seabed terrain database, the operating survey lines of the experimental vessel "Sikuliaq" were mainly distributed in the Chukchi Plateau, as shown in Figure 2. The seabed terrain of the operating area is relatively complex, and the depth of the sea area where the survey lines pass varies from a few hundred to several thousand meters, while the depth of the sea area where the receiving equipment is located is about 1900 m. Therefore, the sound propagation during this experiment will be a complex situation with significant horizontal changes in the seabed terrain. Subsequent research will specifically analyze the impact of terrain changes on the characteristics of acoustic signals.

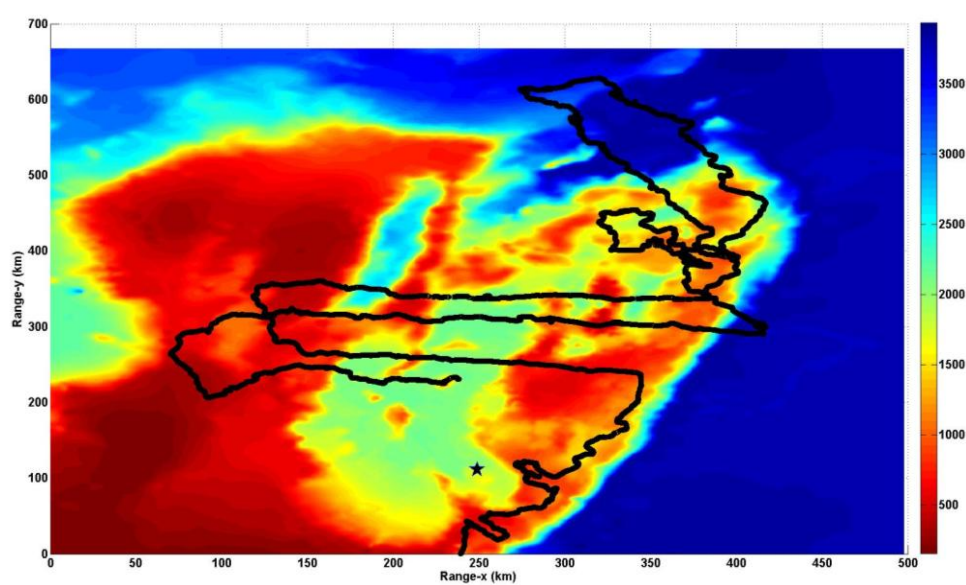

Figure 2 The seabed topography, actual survey line locations, and receiving equipment locations of the seismic survey sea area

## C. Sea ice coverage

The sea ice cover on the Arctic sea surface is one of the important and unique factors that affect the propagation of underwater sound in the Arctic. In order to comprehensively analyze the sound propagation environment in this experiment, this article obtained sea ice cover images of the experimental sea surface observed by satellites during the experiment, as shown in Figure3. Figure3 shows the sea ice concentration with an interval of about ten days. It can be seen from the figure that during the experiment, there was significant sea ice coverage on the surface of the Chukchi Plateau, and the sea ice coverage gradually decreased over the course of about twenty days. However, most of the sound propagation path from the air gun sound source location to the receiving device locations is still covered with sea ice. According to previous research, sea ice coverage has a attenuation effect on the propagation of sound signals.

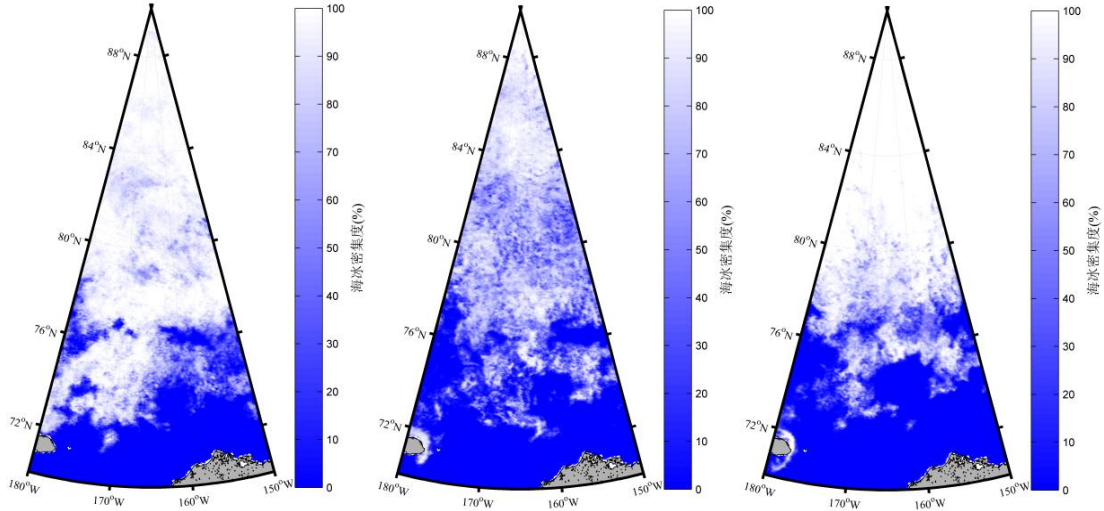

Figure 3 Changes in sea ice cover in the experimental sea area during the experimental period (08/20, 09/01, 09/10)

## D. Sea water sound speed profile

Next, this paper will analyze the most unique SSP in the Chukchi Plateau, which is the dual-channel SSP. This article analyzes the SSP dataset collected from previous experiments in the Chukchi Plateau area, as shown in Figure 4. From this figure, it can be seen that the SSP in the Chukchi Plateau has obvious dual-channel characteristics, but this dual-channel SSP has obvious spatial variation characteristics. Specifically, from the west to the east of the Chukchi Plateau, the dual-channel SSP exhibits a significant change process from scratch to existence, with the critical depth of the upper and lower channels gradually increasing, and the intensity of the lower channels also gradually increasing. Therefore, the sound propagation during this experiment should be a horizontal variation of the dual-channel SSP.

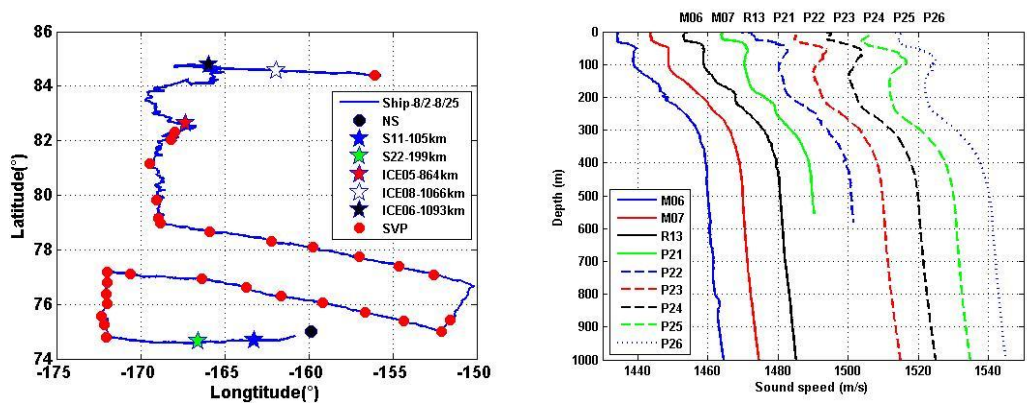

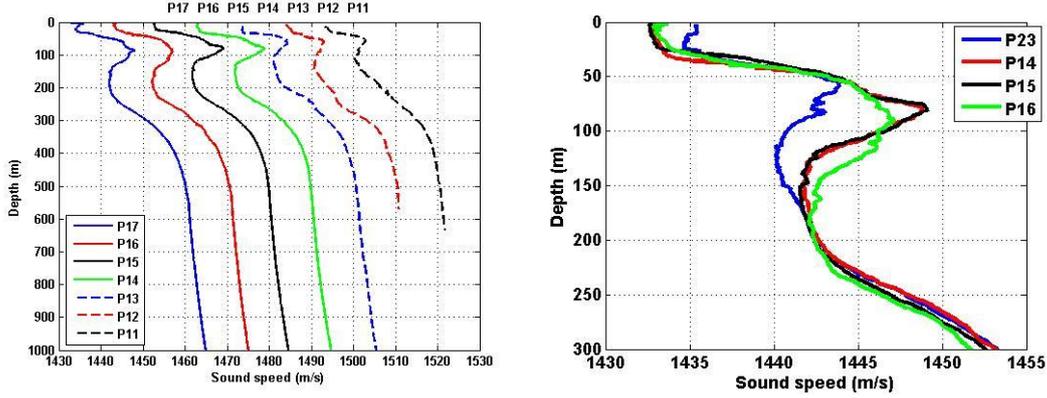

Figure 4 The spatially varying SSP dataset of the Chukchi Plateau measured in previous historical experiments

## IV. SIMULATION OF SOUND FIELD CHARACTERISTICS

Next, this article will conduct sound field feature simulation based on the sound source, reception, and environmental information provided earlier, including calculating the eigenfunctions and dispersion structures based on the normal mode model Kraken, and calculating the sound field transmission loss and sound signal time-domain waveform based on the parabolic equation model Ram.

### A. Simulation of eigenfunctions of normal mode

In order to study the low-frequency dispersion structure of Arctic sound channels, this article first analyzes the vertical distribution of normal mode eigenfunctions at low frequencies. Based on the measured dual-channel SSP, figure 5 shows the vertical distribution of the eigenfunctions of the first three normal modes calculated using the normal mode sound field model Kraken at two frequencies of 20 Hz and 70 Hz.

From this figure, it can be seen that for 20 Hz, the eigenfunction of the first normal mode is mainly distributed within 500 m of the surface, the second normal mode is mainly distributed within 1000 m of the surface, and the third normal mode is mainly distributed within 1500 m of the surface. As the SSP mainly increases with depth, the third normal mode will have a faster group velocity, so at 20 Hz, the third, second, and first normal modes arrive in sequence. For 70 Hz, the energy of the first three normal modes is concentrated within the surface 400 meters, while the water body at the surface 400 meters has a complex dual-channel SSP. Unlike the case at 20 Hz, the eigenfunction of the first normal mode covers more of the relative maximum value of the dual-channel SSP, resulting in the first normal mode having a faster group velocity and arriving earliest.

The above simulation explains the order of arrival of various normal modes in the low-frequency situation. The existence of a dual-channel SSP enables low order normal modes to arrive first and high order normal modes to arrive later in certain frequency band, resulting in a crossover phenomenon on the dispersion curve. Subsequent simulations and experiments will demonstrate this feature.

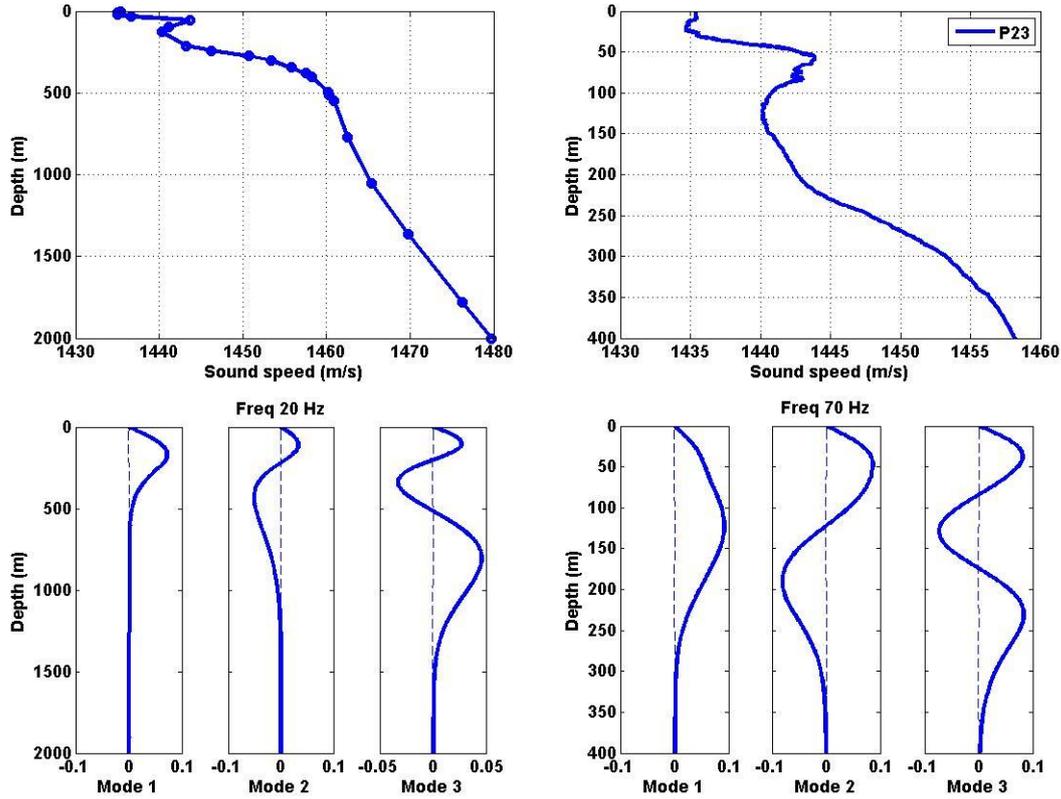

Figure 5 the vertical distribution of the eigenfunctions of the first three normal modes at two frequencies of 20 Hz and 70 Hz based on measured dual-channel SSP

## B. Simulation of two-dimensional sound transmission loss

In order to analyze the impact of the complex seabed topography of the Chukchi Plateau on low-frequency long-distance sound propagation, this paper selects four typical sound source stations with different propagation ranges on the seismic exploration survey line for analysis. The distance from these four stations to the receiving station is 200km, 300km, 400km and 518km respectively.

In the simulation process of sound transmission loss, in order to maintain consistency with the air gun sound source, it is assumed that the sound source depth is 10 m, and the frequency is 100 Hz, and the seabed terrain is calculated using the ETOP database. This article uses the parabolic equation model RAM for calculation. Figure 6 shows the two-dimensional sound transmission loss of the four simulated stations corresponding to different seabed terrain conditions.

From this figure, it can be seen that for the case of 200km, the seabed terrain does not change much, and the sea depth varies between 1500 m and 2000 m. Combined with the vertical distribution of the eigenfunctions in figure 5, the seabed terrain basically does not block the low-frequency part of the normal mode eigenfunction, and the receiving end has sound energy reaching both medium and large depths. For the case of 518 km, the sea depth changes from 3500 m to 1500 m, and the seabed terrain has less obstruction to the low-frequency part of the normal mode, and sound energy can be received at medium and large depths. For the cases of 300 km and 400 m, there are obvious seamounts on the propagation path, which have obscured the low-frequency eigenfunctions of the normal mode, and there is almost no sound energy reaching at medium and large depths. Therefore, the propagation of sound in the deep Arctic sea is greatly influenced by the seabed terrain, especially in cases of large reception depths. The complex seabed terrain can block the long-distance propagation of the low-frequency part of the normal mode, resulting in the absence of sound energy at medium and large depths. Subsequent experimental data also supports this viewpoint.

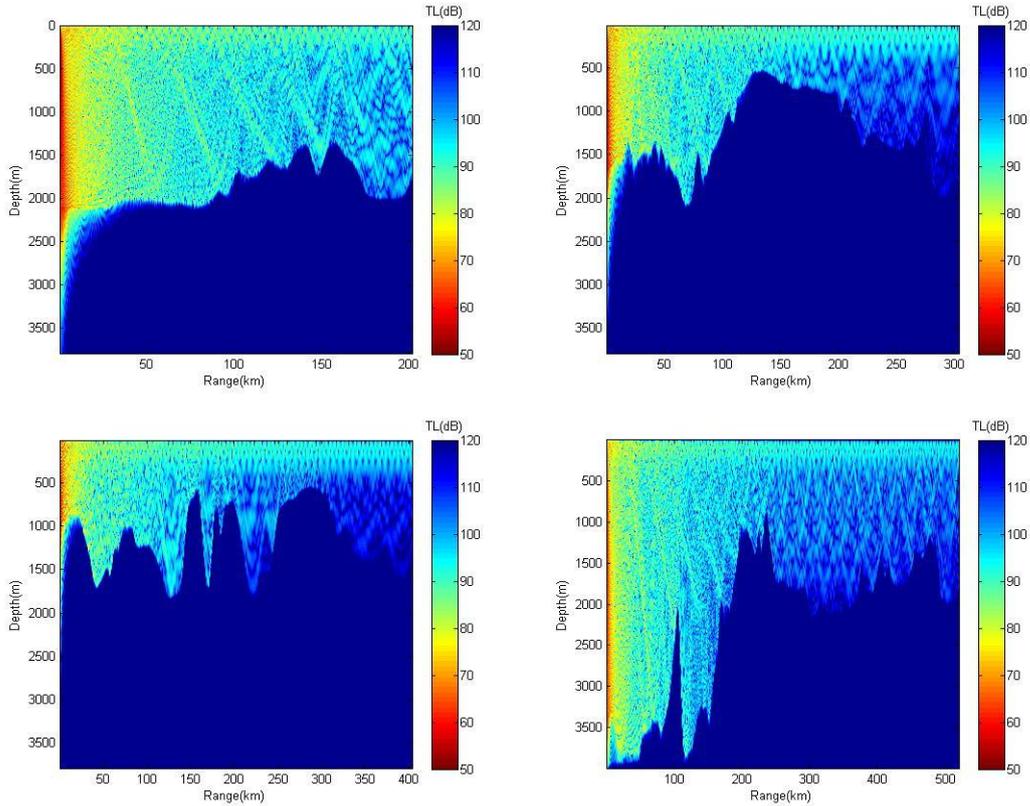

Figure 6 Simulation results of two-dimensional sound transmission loss from four typical sound source stations with different propagation distances to receiving stations

## C. Time-domain waveform simulation of received acoustic signals at different propagation ranges

Next, this paper conducts time-domain wave simulation of the received acoustic signals at the four propagation distances mentioned above, and provides corresponding signal time-frequency analysis diagrams. In order to analyze the dispersion structure, this paper calculates the group velocity of normal modes based on Kraken to obtain the dispersion structure. However, unlike acoustic signal simulation, the dispersion structure simulation uses an ocean environment with unchanged seabed terrain and a fixed sea depth of 2000 m for simulation. Finally, this article compares and analyzes the dispersion curve obtained from simulation with the time-frequency analysis of the simulated sound signal.

From the time-domain waveform and time-frequency analysis of simulated acoustic signals in the figure 7, it can be seen that both 200km and 518km acoustic signals have complete normal mode arrival structures, with the low-frequency part of the normal mode covering a large depth range arriving at the beginning, which is basically not affected by seamounts, and the high-frequency part of the normal mode covering the surface seawater arriving at the end. The sound signals at 300km and 400km are incomplete normal mode arrival structures without a clear signal beginning, meaning that the low-frequency part of the normal mode wave is blocked by the seamount and cannot be reached, while the high-frequency part is not affected.

From the figure 7, it can be seen that the simulated dispersion curve is basically consistent with the dispersion structure of the simulated acoustic signal, indicating that in the case of complex seabed terrain, the dispersion structure simulation calculation based on the normal mode model of the range-independent seabed can be used, without the need for calculations based on segmented approximation and large computational complexity of the adiabatic normal mode model. The simulation comparison between 200km and 518km shows that the simulated acoustic signals of both have a complete dispersion structure. The simulation comparison between 300km and 400km shows that they do not have a complete dispersion structure, and the low-frequency part has been obscured by the seamounts.

In addition, the dispersion structure of the simulated acoustic signal and the dispersion curve of the simulation are consistent with the previous analysis combined with the sound velocity profile and eigenfunction. In the low-frequency part (such as the 20Hz part), the first normal mode arrives the latest; while in the high-frequency part (such as 70Hz), the first normal mode arrives the earliest.

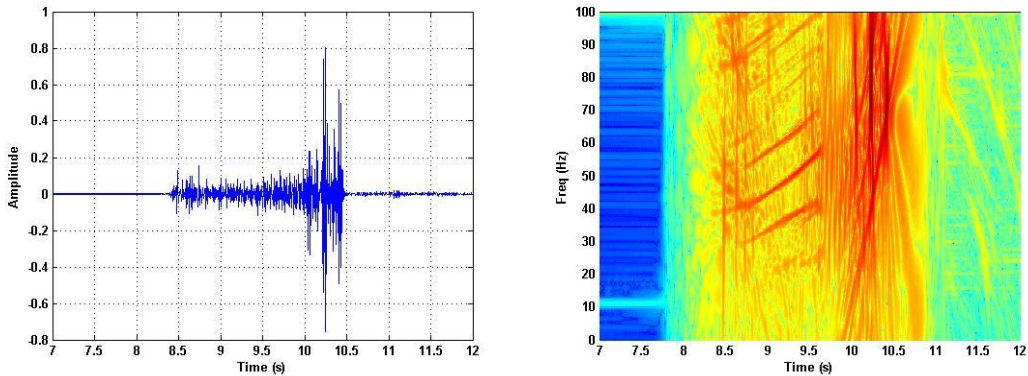

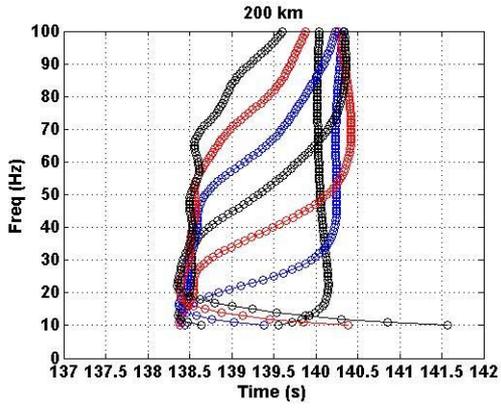 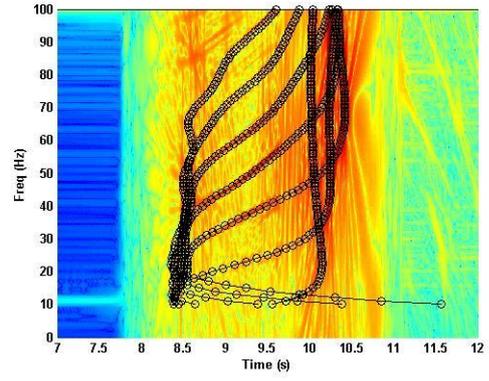

(a) 200 km

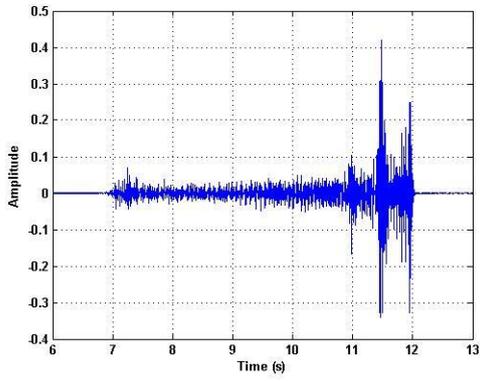 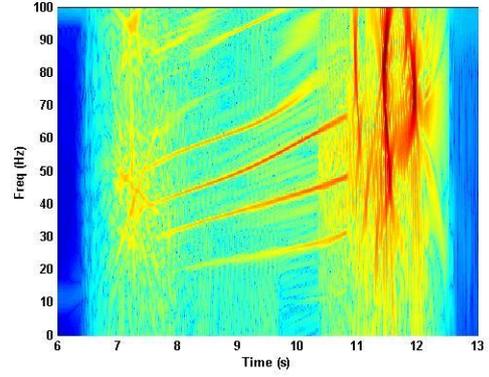

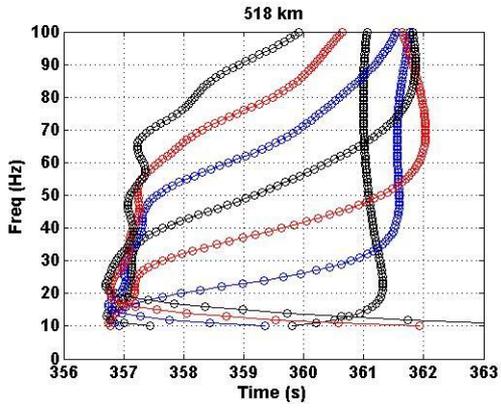 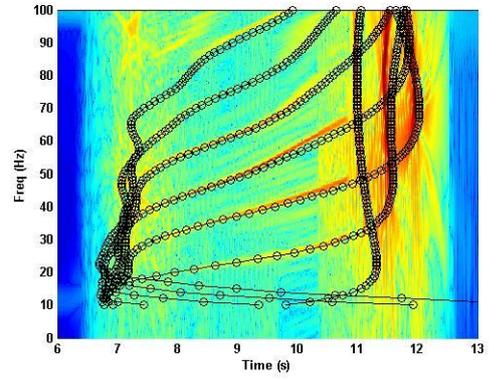

(b) 518 km

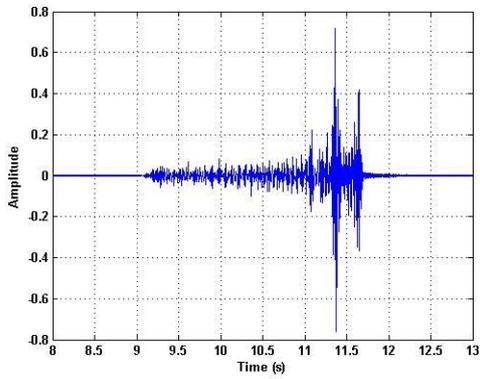 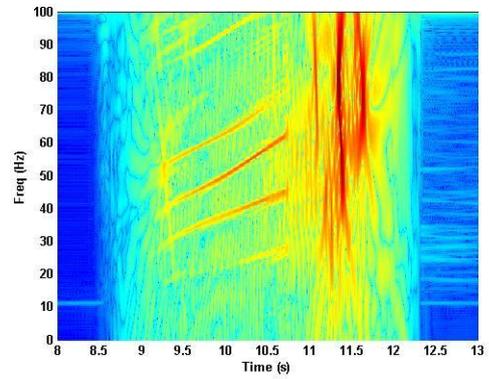

**11**

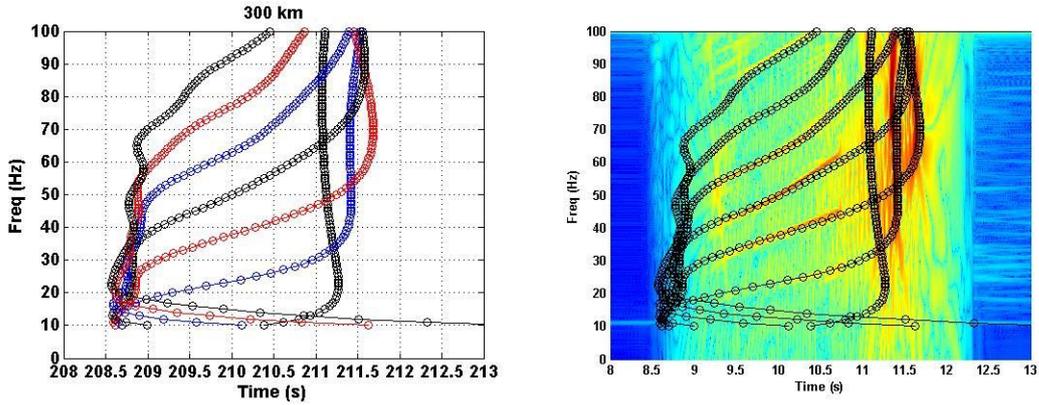

(c) 300 km

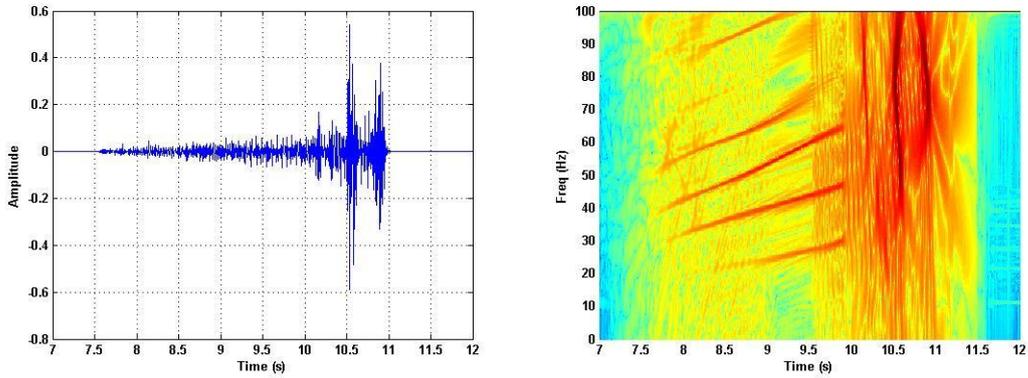

(d) 400 km

Figure 7 Time domain waveform and time-frequency analysis of sound signals based on RAM at four propagation distances, and dispersion curve based on Kraken calculation

## D. Time-domain waveform simulation of received acoustic signals at medium and large depths

The time-domain waveform and time-frequency analysis of acoustic signals with medium and large reception depths are shown in the figure 8. For the cases of 300 km and 400 km, due to the obstruction of seamounts, there is basically no sound energy reaching at these two receiving depths, so this situation is ignored and not analyzed.

For medium and large depths of 200km and 518km, it can be seen from the figure that the time-domain waveform follows a series of gradually decreasing peaks after the maximum peak. It can be understood that the maximum peak is the sound ray that has not been reflected by the seabed and is reversed at large depths, with the shortest path and the smallest attenuation of the seabed reflection, resulting in the fastest arrival and maximum amplitude. Subsequently, the sound signal that has been reflected multiple times by the seabed arrives. Therefore, the acoustic signals at medium and large depths mainly exhibit the properties of ray acoustics, and the signals do not have obvious dispersion structures. When the seabed attenuation is large, only the maximum and fastest peak can be observed in actual experiments.

The time-domain waveform and dispersion characteristics of sound signals at medium and large depths are completely inconsistent with the received sound signals at the surface. In the surface water, the maximum peak of the surface sound signal arrives at the latest, and this part of the signal is not affected by the seabed; while the starting part of the signal is attenuated due to the influence of the seabed. The entire signal exhibits the property of a normal mode, and has a clear dispersion structure. Therefore, the sound propagation in the deep Arctic has completely different sound propagation properties in surface, middle, and bottom water bodies.

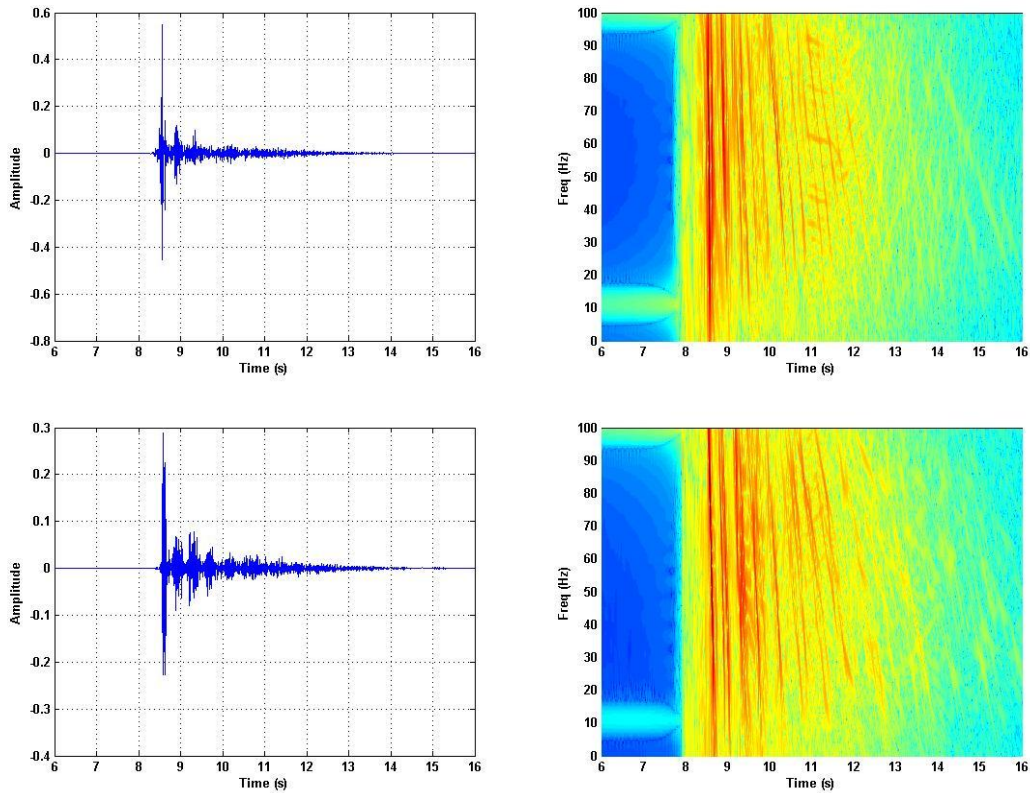

(a) At the depth of 883 m and 1662 m with the propagation range of 200 km

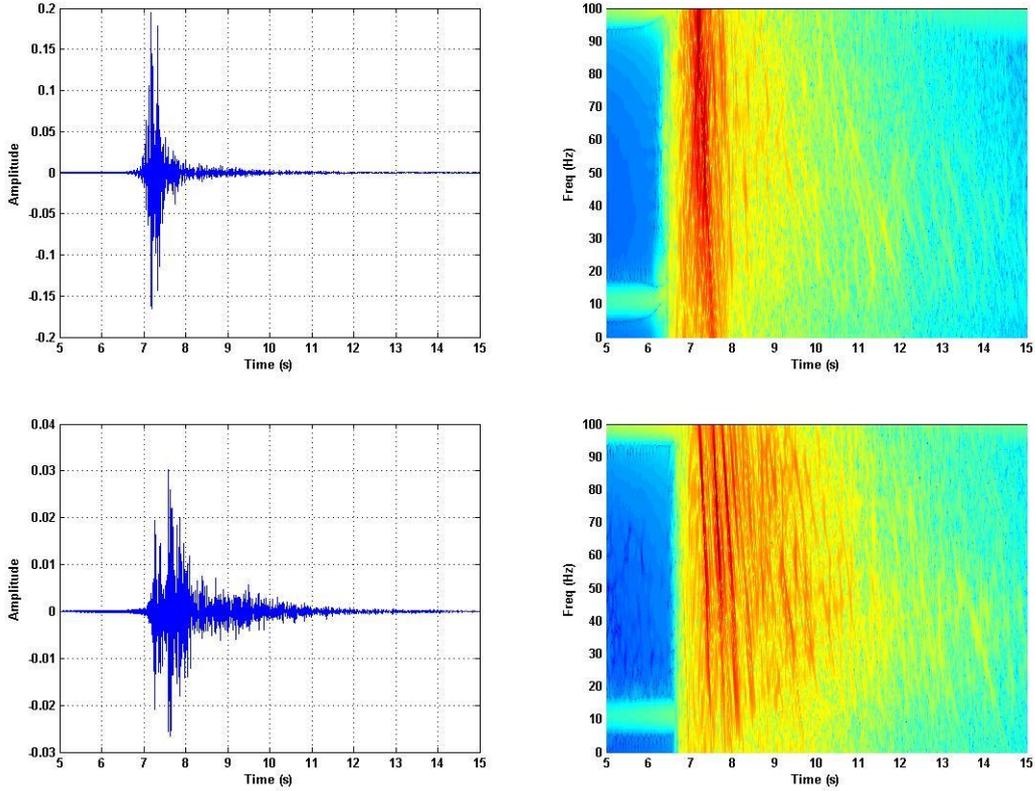

(b) At the depth of 883 m and 1662 m with the propagation range of 518 km

Figure 8 The time-domain waveform and time-frequency analysis of received acoustic signals at medium and large depths at propagation distances of 200 km and 518 km

## V. EXPERIMENTAL DATA PROCESSING

Next, this article will conduct experimental data analysis and verify the previously mentioned viewpoints by comparing them with simulation results, including the dispersion structure characteristics of acoustic signals received at different distances at surface depths, the characteristics of acoustic signals received at different depths, and the modal separation work of received acoustic signals.

### A. Time-domain waveform and dispersion structure of air gun acoustic signal at different propagation ranges

Figure 9 shows the time-domain waveform and time-frequency analysis of the received signals at 200 km and 518 km. From this figure, it can be seen that the time-domain waveform and dispersion structure of the acoustic signal have consistent patterns, and there two signals have a complete normal mode dispersion structure.

For the 200km signal, it can be seen from the time-domain signal graph and time-frequency analysis graph that the experimental received signal has a clear and complete dispersion structure of normal modes. The first pulse signal to arrive is the lowest frequency part of each normal mode, forming a clear pulse peak. The last pulse signal to arrive is the high-frequency part of each normal mode, with a time difference of about 1.5 seconds. Among them, in the low-frequency part, the first normal mode significantly lags behind the other normal modes and has the minimum group velocity; while in the high-frequency part, the first normal mode has the maximum group velocity and arrives the earliest. Therefore, there is a clear intersection of the dispersion curves of the normal mode in the high-frequency part.

For the 518km signal, it can be seen from the time-domain waveform and time-frequency analysis that it also has a complete dispersion structure of normal mode, and the time difference between the first arriving pulse signal and the last arriving pulse signal is about 4 seconds, which is significantly larger than the case of 200km, and is approximately proportional to the distance relationship.

For the cases of 200 km and 500 km, this article compares the dispersion structure curve of the sound field simulation with the experimental dispersion structure. At 200 km, the two are basically consistent, including the duration of the entire dispersion structure and the high-frequency part of the first three normal modes. For the case of 518km, the two are basically consistent. However, there are some differences between the two, including the dispersion structure of the middle and high-frequency parts. Considering that the sound velocity profile of the Chukchi Plateau in the experimental sea area varies significantly with space, especially for the case of 518km, when the propagation distance is far, the horizontal variation of the sound velocity profile needs to be considered, which was not taken into account in the simulation, resulting in inconsistency between the two.

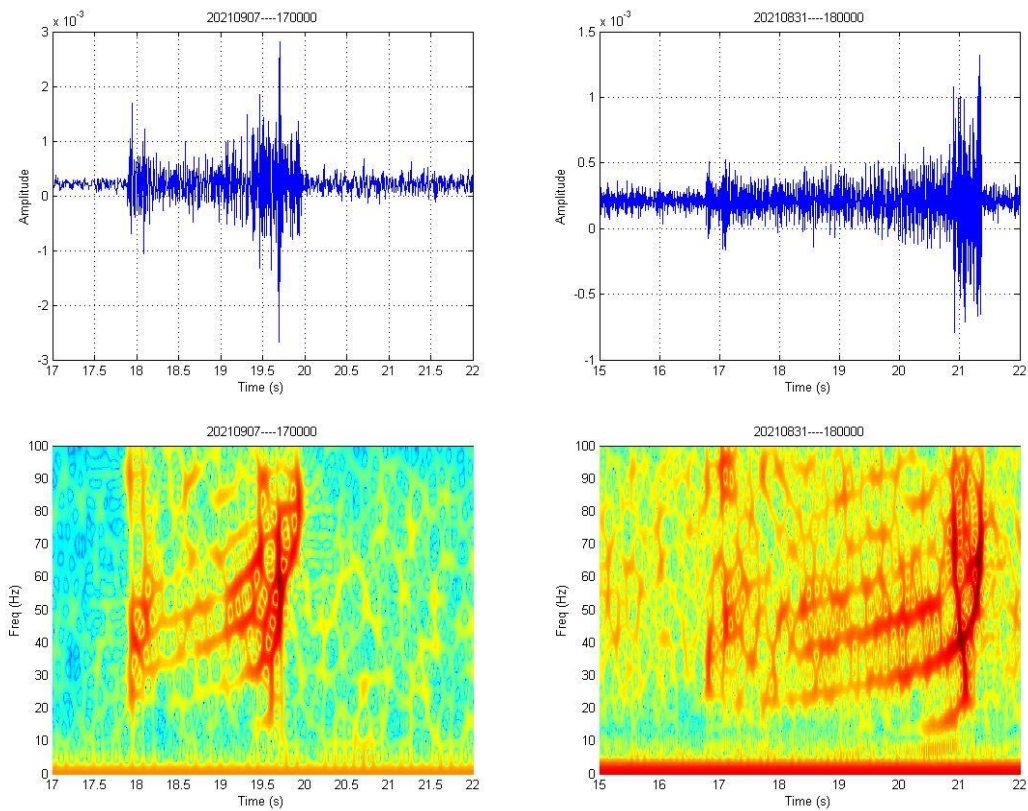

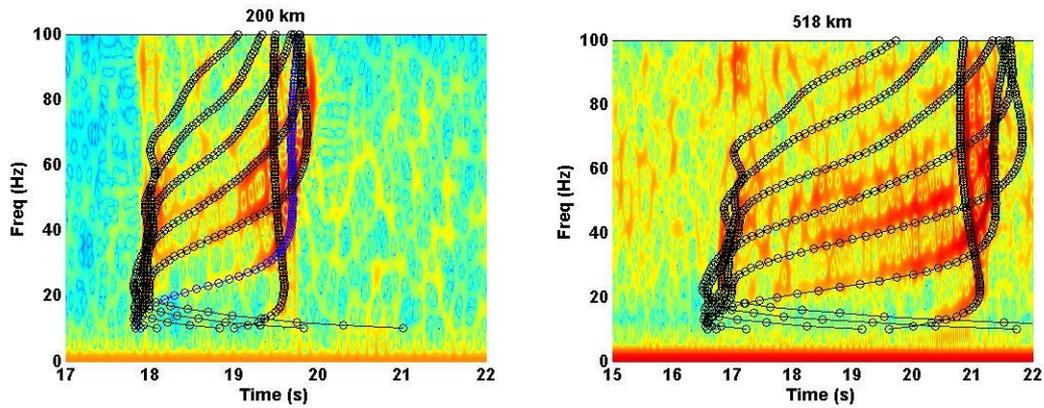

Figure 9 The time-domain waveform and time-frequency analysis of received acoustic signals with propagation distances of 200 km and 518 km, as well as comparison with simulated dispersion curves

For the cases of 300km and 400km, Figure 10 shows the time-domain waveform and time-frequency analysis of the experimentally collected acoustic signal. It can be seen that there is no obvious signal starting time in the time-domain signal, mainly the low-frequency part of the normal mode wave is obstructed by the seamount, which is consistent with the simulation results. On the time-frequency analysis chart, strong first four normal modes can still be clearly seen. In the low-frequency part, the first normal mode is still the last to arrive, and in the high-frequency part, the first normal mode is still the earliest to arrive, and this figure has a clear third normal mode. But at 400km, the high-frequency part of the third normal mode has significantly decayed, and the high-frequency part of the third normal mode has basically disappeared at 518km.

Comparing the dispersion curve of the sound field simulation calculation with the dispersion structure of the experimental measured sound signal, as shown in the figure, it can be seen that the low-frequency part of the experimental signal has disappeared due to the influence of the seamount. In the middle part, the two are basically the same. In the high-frequency part, the dispersion structure of the first three numbers is also basically the same. The above four signals describe the terrain effect very well.

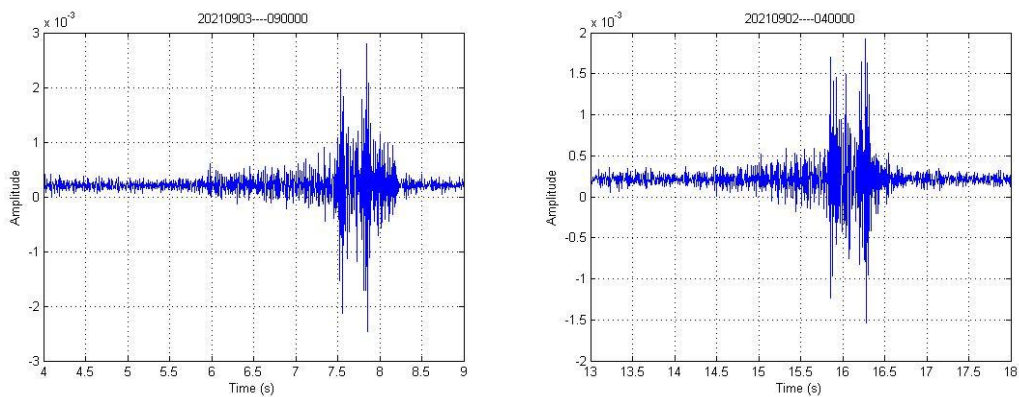

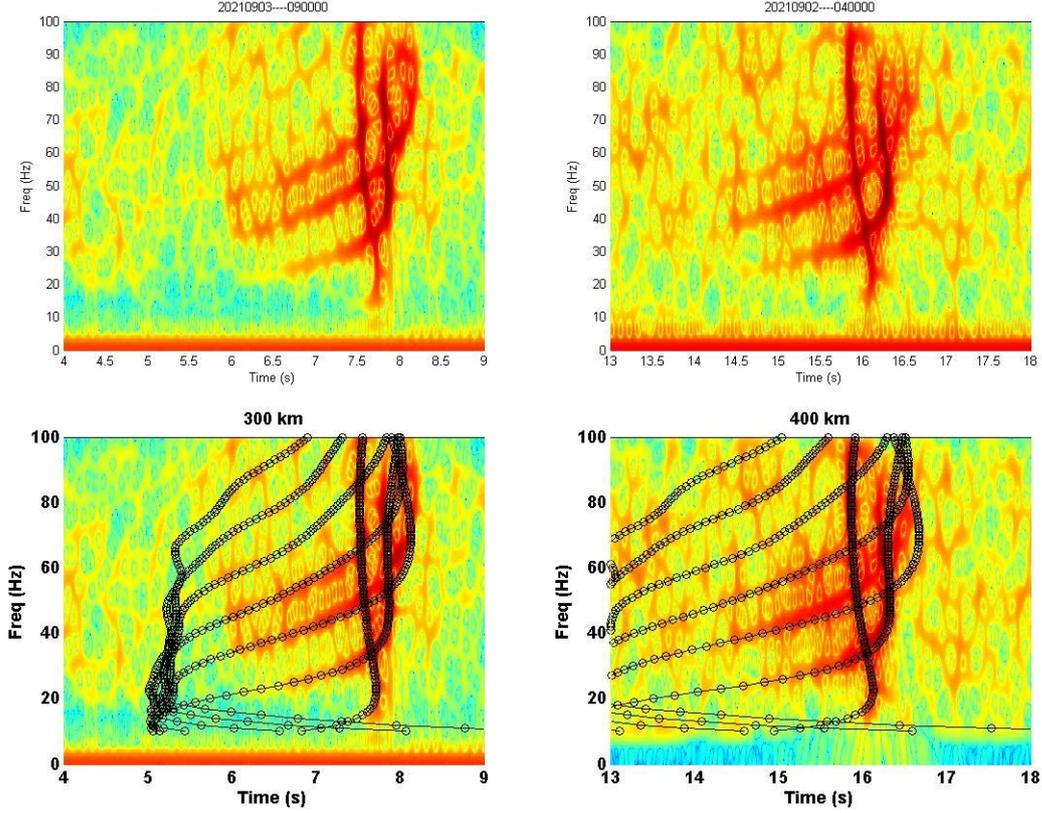

Figure 10 The time-domain waveform and time-frequency analysis of received acoustic signals with propagation distances of 300 km and 400 km, as well as comparison with simulated dispersion curves

## B. Warping transformation and normal mode separation of air gun acoustic signal

The following is a warping transform based method for separating normal modes and extracting dispersion structures of air gun signals. Considering the characteristics of the acoustic signal from the Chukchi Plateau, the intermediate section of the acoustic signal with complete dispersion structure is intercepted for warping transform of refractive normal modes, avoiding the intersection of Airy phase points and dispersion curves, as shown in the following figure 11.

For the case of 200km, as shown in the figure, the duration of the dispersion section in the middle of the acoustic signal is about 1.5s. The middle section is cut off for analysis. From the time-frequency analysis graph, it can be seen that there are obvious second, third, fourth, and fifth normal modes, while the sixth and seventh normal modes have significant attenuation and low signal-to-noise ratio. Through refractive warping transformation, spectral analysis and time-frequency analysis can be performed, and six obvious approximate single frequency signals can be seen, Namely, the second to seventh normal modes, among which the energy of the second to fifth modes is relatively high. After band-pass filtering and inverse warping transformation, the time-domain waveform of the four normal modes can be obtained, as shown in the figure. The

**17**

amplitude of the second normal mode is relatively small. Afterwards, the dispersion curve of a single mode of the separated four modes is extracted, and the dispersion curve of the entire acoustic signal is finally obtained. Compared with the time-frequency analysis diagram of the measured signal, the two are consistent, which indicated that using the warping transformation of refractive mode can effectively separate the mode and extract the dispersion structure of broadband pulse acoustic signals in the deep Arctic sea.

For the 518km signal, the middle section signal in complete dispersion structure is also intercepted, avoiding the intersection of Airy phase points and dispersion curves. After refractive warping transformation, spectral analysis and time-frequency analysis can be performed, and six obvious approximate single frequency signals can be seen, representing the first to sixth normal modes. After band-pass filtering and inverse warping transformation, strong first four normal modes can be separated, and the modal dispersion curve can be extracted. Compared with the time-frequency analysis of the measured signal, the two are consistent.

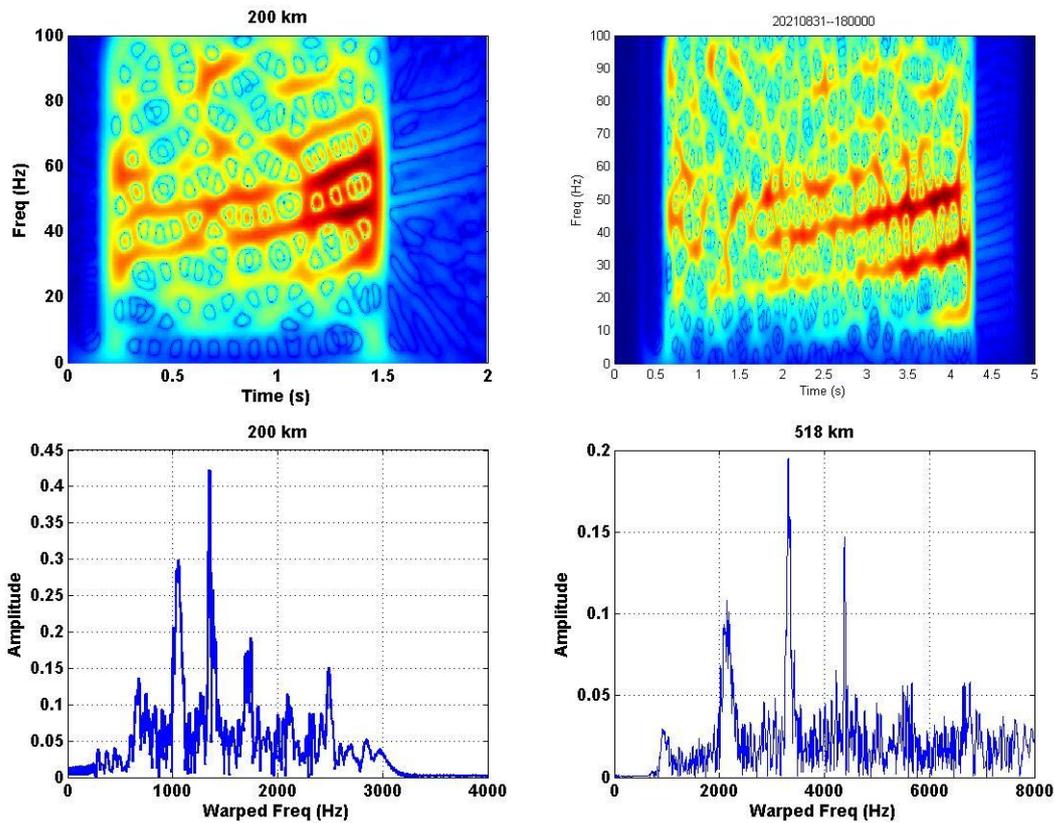

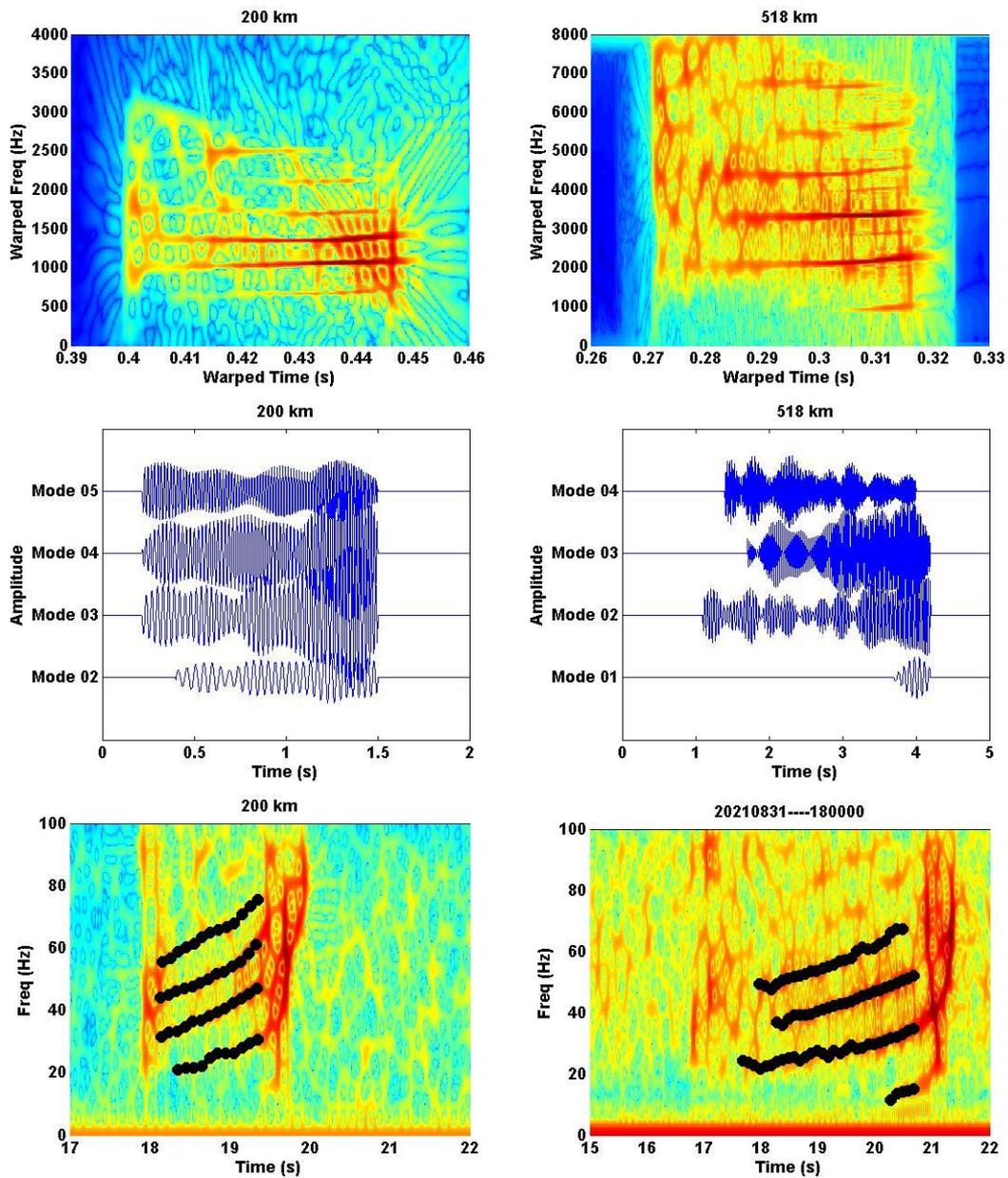

Figure 11 The process of normal mode separation and dispersion curve extraction based on warping transform for received acoustic signals with propagation distances of 200 km and 518 km

For signals at 300km and 400km, as there is no obvious signal starting time and there is no sound signal with a complete dispersion structure, the signal should be intercepted avoiding the intersection of the dispersion curve for warping transform.

For a 300km signal, approximately 1.5s of the acoustic signal is intercepted for transformation. After spectral analysis and time-frequency analysis, obvious second, third, and fourth modes can be seen, as well as weaker fifth, sixth, and seventh modes. After band-pass filtering and inverse warping transformation, the time-domain waveforms and dispersion curves of the second to fifth modes can be obtained, which are compared with the dispersion structure of the measured acoustic signal, the two are in agreement.

For a signal of 400km, an acoustic signal of about 1.5s is intercepted for transformation. After transformation, obvious second, third, fourth, and sixth mode waves can be observed, and the fifth mode wave is basically not visible. Due to the sound source depth of about 10m and not located at the nodes of each mode wave, the reception depth may be located near the nodes of the fifth mode wave, so the fifth mode wave cannot be received. After inverse warping transformation, the time-domain signals and dispersion curves of each mode are obtained, which are consistent with the dispersion structure of the measured signal, as shown in the figure 12.

The refractive warping transformation of these four typical signals above indicates that the broadband pulse acoustic signals from the Chukchi Plateau in the Arctic can be subjected to warping transform based on refractive normal modes, separating each normal mode and extracting dispersion structures. This can be directly used for marine environment inversion and sound source localization research based on dispersion structures. And those work will be carried out in the future.

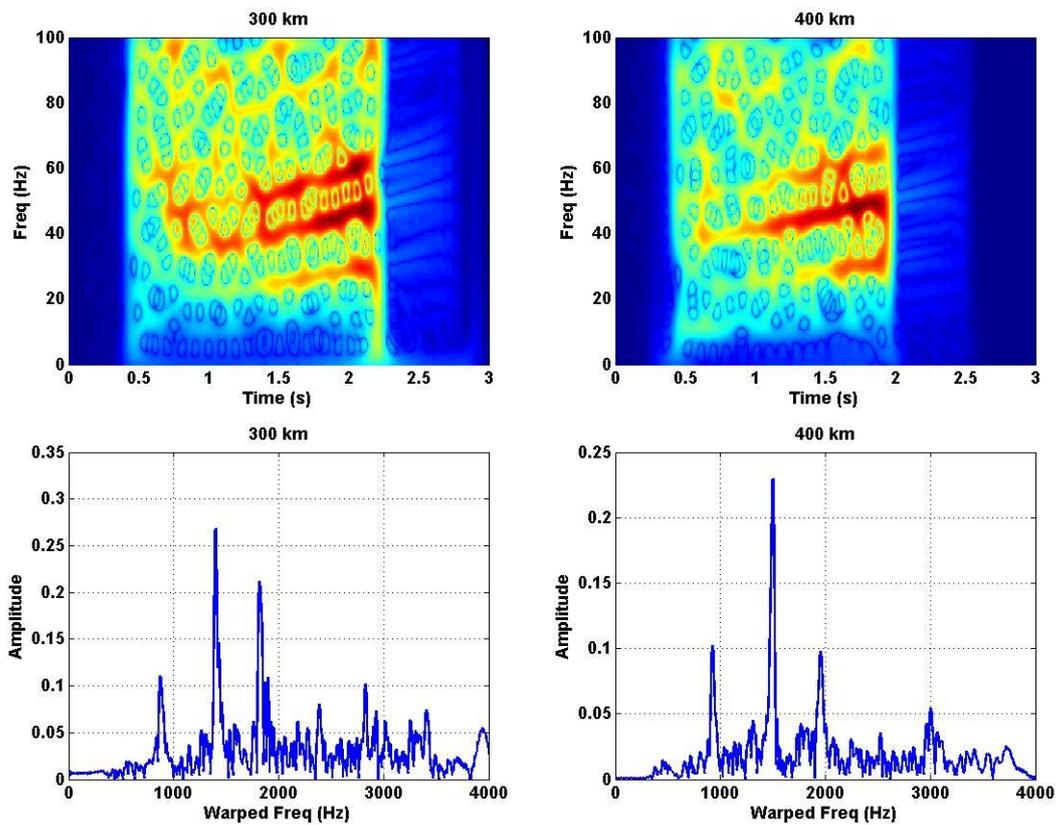

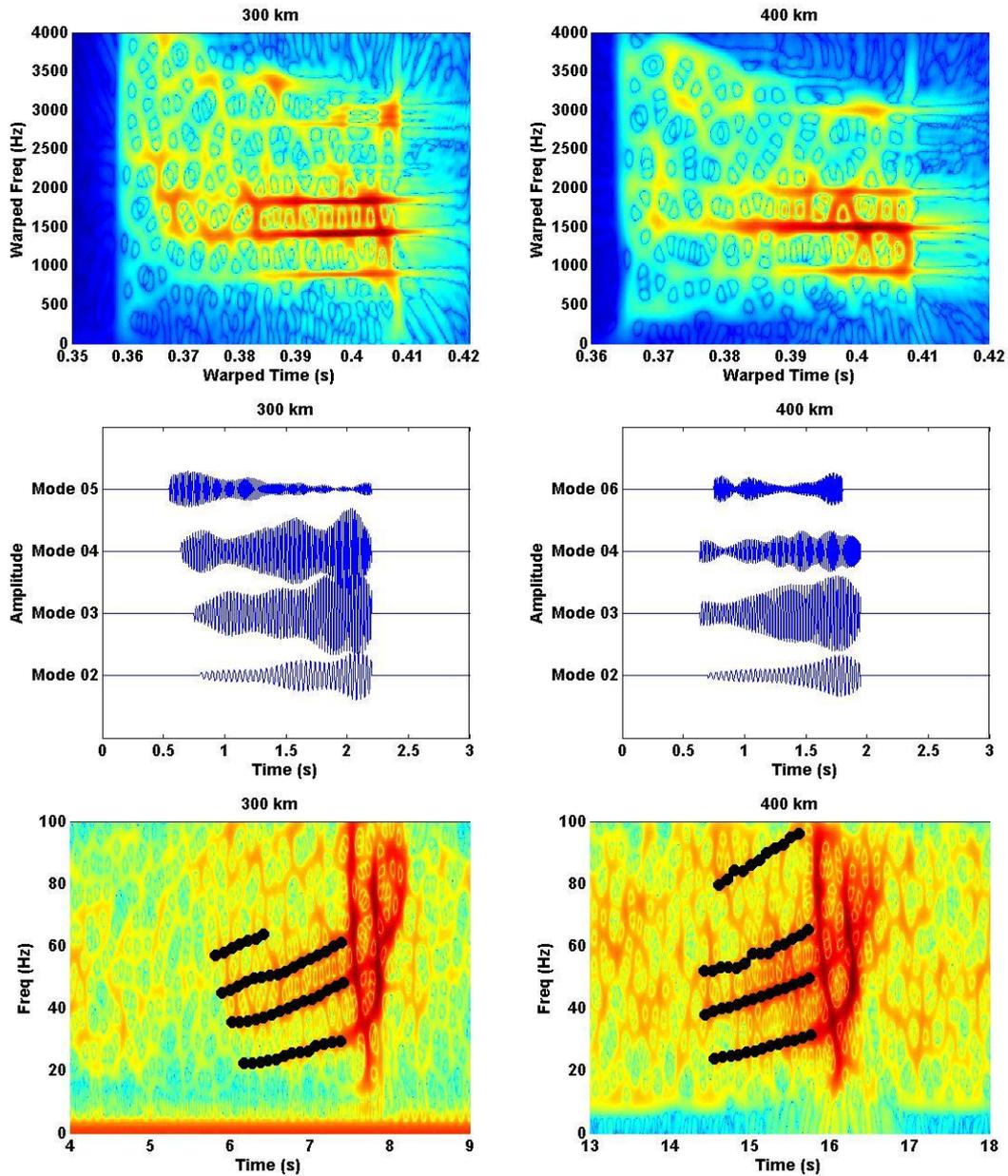

Figure 12 The process of normal mode separation and dispersion curve extraction based on warping transform for received acoustic signals with propagation distances of 300 km and 400 km

## C. Time-domain waveform and dispersion structure of air gun acoustic signal at medium and large depths

The following is an analysis of the received acoustic signals for experiments at medium and large depths. Figure 13 shows the time-domain waveform and time-frequency analysis of the acoustic signals at two depths at distances of 200 km and 518 km. For the case of 200 km, it can be seen from the figure that the characteristic of the time-domain of the acoustic signal is that the maximum peak first arrives, and then the signal amplitude gradually decreases. The acoustic signal does not have a clear dispersion structure, which is basically consistent with the analysis in the simulation. The acoustic signal received at a large depth exhibits multipath acoustic ray properties, and the acoustic signal gradually decreases with the increase of the number of reflections on the seabed. For the case of 518 km, only the initial peak arrival signal can be observed in the experimental data, and the subsequent acoustic signals reflected multiple times on the seabed no longer have a signal-to-noise ratio and cannot be observed, which is consistent with the conclusion in the simulation.

For the medium and large depth receiving equipment at 300 km and 400 km, due to the obstruction of underwater mountains, the acoustic signal is submerged by noise, and no obvious signal can be found in the data, which is basically consistent with the previous simulation analysis. During the sound propagation of deep water in the Arctic, the bottom mountains have a strong attenuation effect on the reception of sound signals at large depths, which is not conducive to sound signal detection.

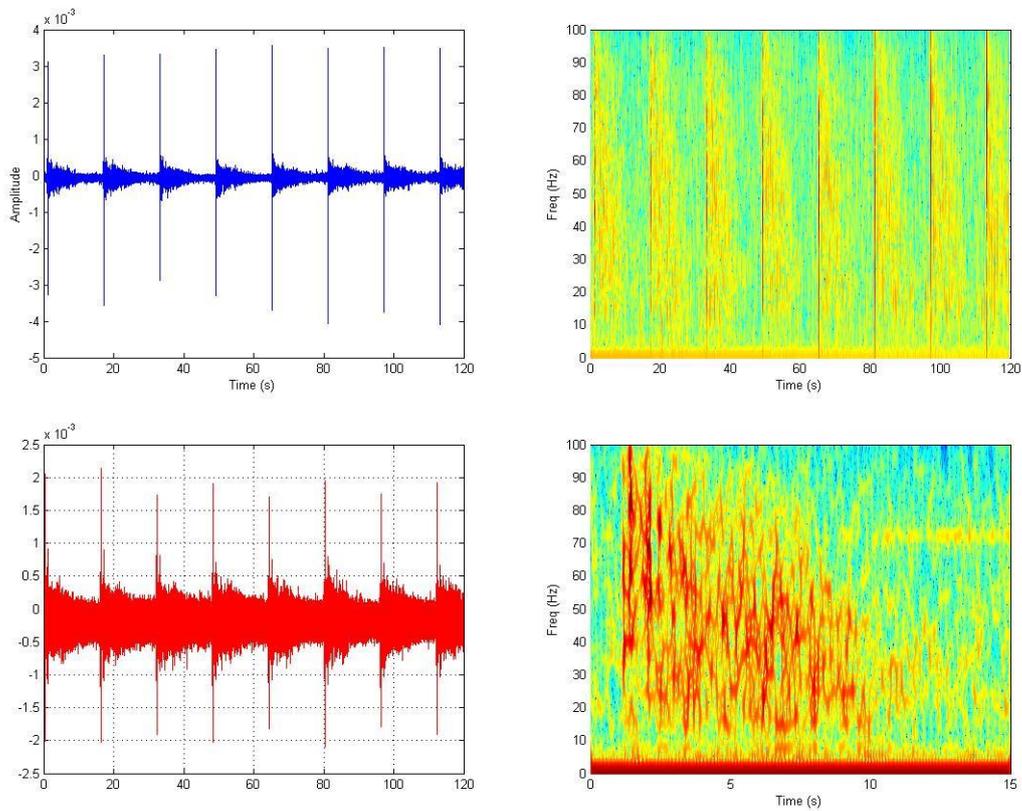

(a) 200 km 700m and 1730 m

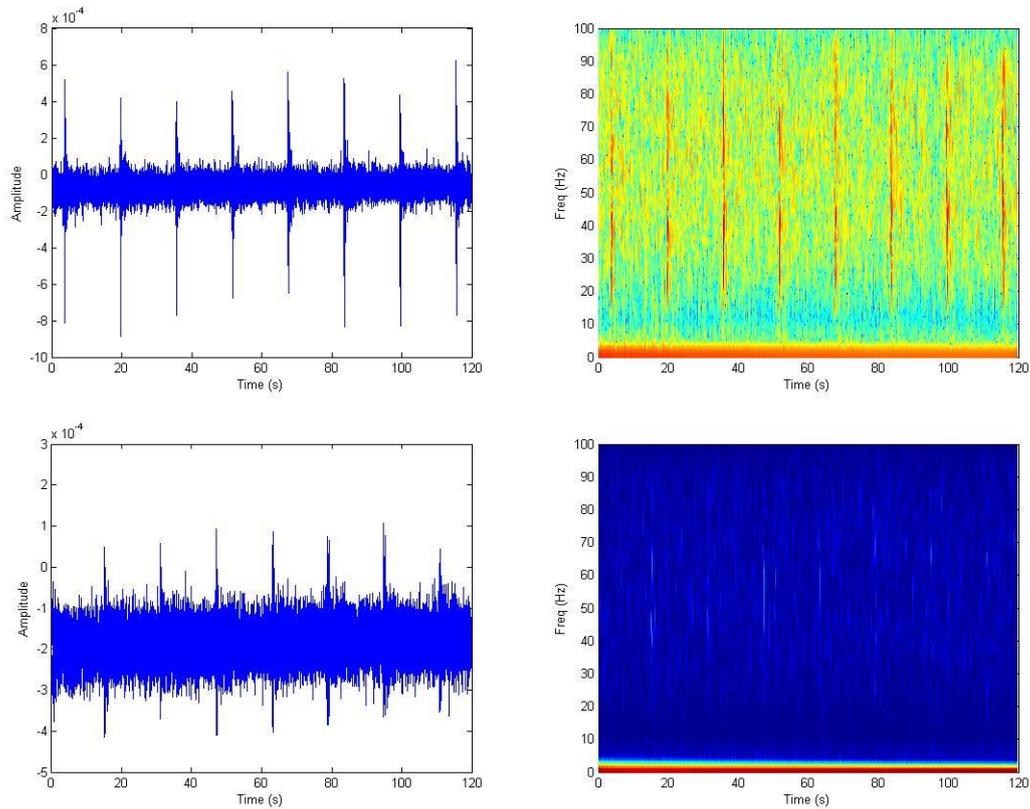

(b) 518 km 700 m and 1730m

Figure 13 The time-domain waveform and time-frequency analysis of air gun sound signals at different reception depths (700m and 1730m) with propagation ranges of 200km and 518km

# VI. TWO METHODS FOR ESTIMATING SOUND SOURCE RANGE BASED ON TIME-DOMAIN WAVEFORMS OF ACOUSTIC SIGNALS UNDER A DUAL-CHANNEL SSP

From the previous analysis of the characteristics of acoustic signals, it can be seen that the time-domain waveform structure is closely related to the distance from the sound source, including the entire signal duration (i.e. the duration of the dispersion structure) and the partial pulse arrival time difference (i.e. the difference in arrival time of the normal mode). Both of these times increase with the increase of propagation distance. Therefore, this paper proposes two fast estimation methods for the distance from the sound source based on the time-domain waveform of the acoustic signal.

## A. A method of estimating sound source range based on the duration of normal mode dispersion structure

This method uses the time difference between the starting and ending peaks of the signal to estimate the sound source distance. Among them, the arrival time of the

**23**

initial peak can be approximated as the distance from the sound source divided by the average sound velocity of the entire sea depth, and the arrival time of the end peak can be approximated as the distance from the sound source divided by the average sound velocity of the surface seawater. This method is relatively insensitive to changes in the sound velocity profile of Arctic seawater and has good stability in the results.

Firstly, as shown in Figure 14, this article obtains the signal time-domain envelope based on the original acoustic signal waveform, and then combines the signal envelope and time-frequency analysis to determine the peak arrival time to obtain the arrival time difference. For a signal of 200 km, the starting time is the arrival time of the second peak, which is 18.0919s; the ending time is the arrival time of the penultimate peak, which is 19.6958s; the duration of the dispersion structure of the normal mode is 1.6039 seconds. For a signal of 518 km, the starting time is the arrival time of the second peak, which is 17.1042s; the ending time is the arrival time of the last peak, which is 21.3239s; the duration of the dispersion structure of the normal mode is 4.2197s.

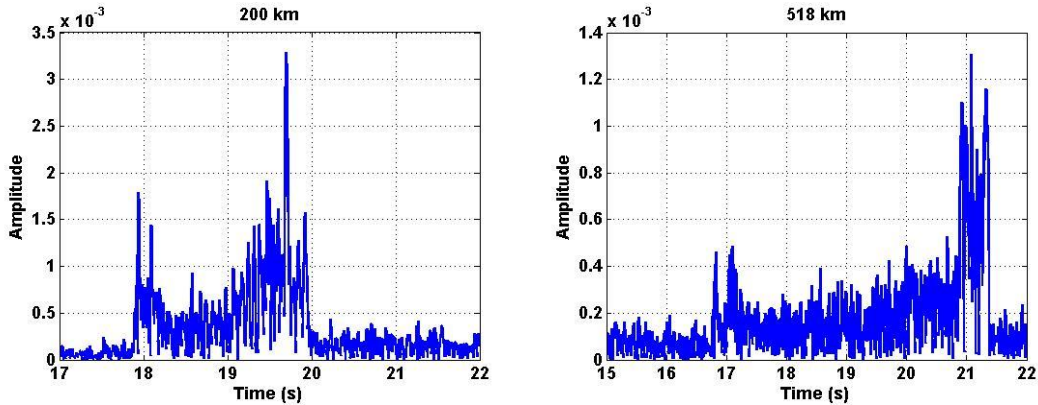

Figure 14 Time domain waveform envelope of acoustic signals with propagation distances of 200km and 518km

Afterwards, as shown in Figure 15, based on the measured sound velocity profile, this article uses the Kraken model to calculate the group velocity of the top ten normal modes within 100Hz, and obtains the group velocity dispersion curve. From the figure, it can be seen that except for the first normal wave, the duration of the remaining normal waves is basically the same. Therefore, based on the group velocity dispersion curves of each normal wave, equivalent average starting and ending group velocities are obtained. Finally, as shown in Figure 16, based on the estimated duration of the dispersion structure of the measured signal and the equivalent average group velocity obtained from model simulation, the estimated sound source distance of the 200km sound signal is 192.63km, and the estimated sound source distance of the 518km sound signal is 506.80km. The estimated sound source distance is relatively accurate and has small errors.

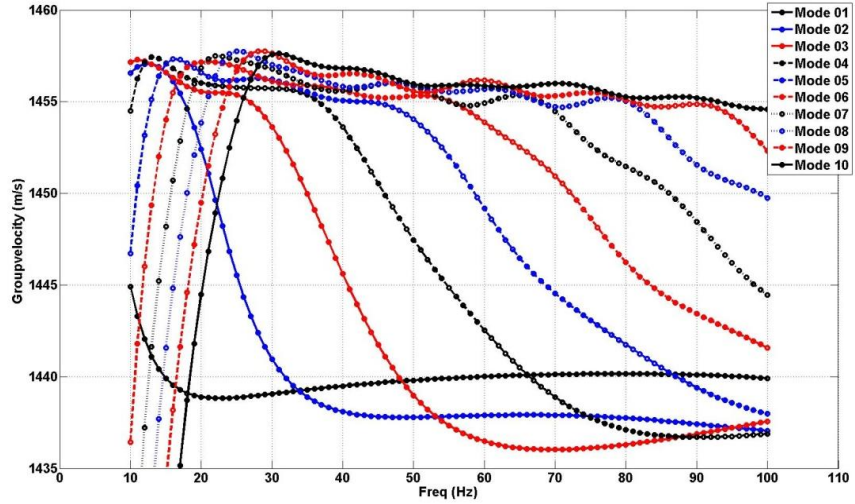

Figure 15 Simulation of Normal Mode Group Velocity Dispersion Curve Based on Measured Marine Environmental Parameters

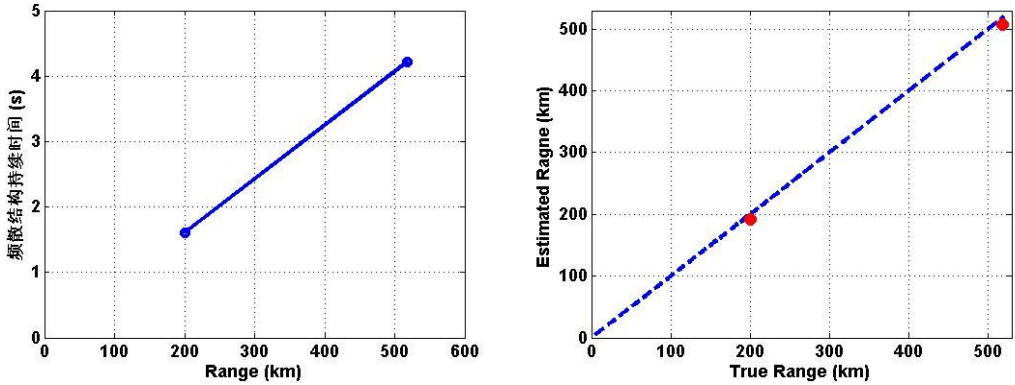

Figure 16 Estimation of dispersion structure duration and corresponding sound source distance based on measured signal estimation

## B. A method of estimating sound source range based on arrival time difference of normal modes

This method uses the time difference of arrival between the first and second modes to estimate the sound source distance. Among them, the time difference of arrival between the first and second modes is extracted using the signal envelope of experimental data, and the equivalent group velocities of the first and second modes are calculated using the Kraken model based on the measured marine environment. The estimation of the sound source distance is obtained by combining these two. Due to the fact that the received acoustic signal is a broadband acoustic signal, it is necessary to calculate the equivalent group velocity within the acoustic signal bandwidth. In addition, this method is achieved by utilizing the arrival time difference between the first and second normal modes in a dual-channel SSP environment. Therefore, this method is more sensitive to surface dual-channel SSP and cannot be applied in the absence of dual-channel SSP.

Firstly, as shown in Figure 17, this article obtains the time-domain envelope of the signal based on the original acoustic signal waveform, and then combines the envelope of the signal with time-frequency analysis to determine the arrival time of the normal mode wave to obtain the difference in arrival time. For a signal of 300km, the arrival time of the first normal mode is the arrival time of the second to last peak, which is 7.5111 seconds; the arrival time of the second normal wave is the arrival time of the penultimate peak, which is 7.8536 seconds; the difference in arrival time between the first and second normal modes is 0.3425 seconds. For a signal of 400km, the arrival time of the first normal mode is the arrival time of the penultimate peak, which is 15.8549s; the arrival time of the second normal wave is the arrival time of the penultimate peak, which is 16.2747 seconds; the difference in arrival time between the first and second normal modes is 0.4198s.

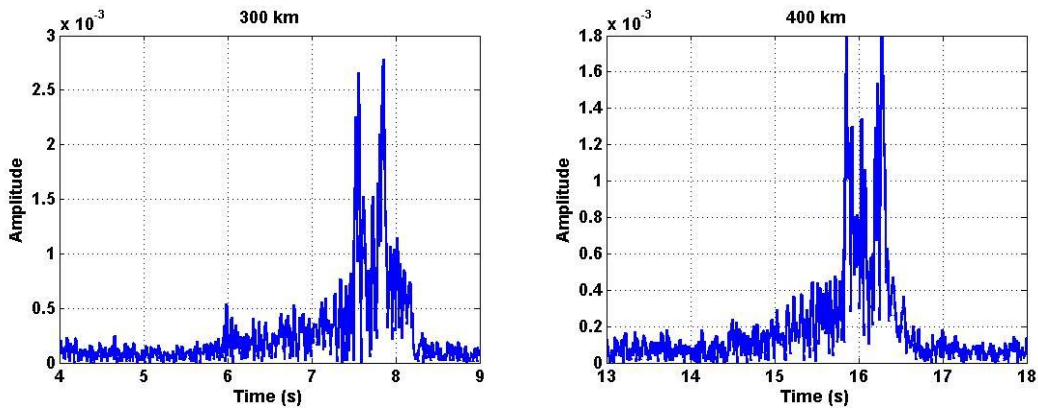

Figure 17 Time domain waveform envelope of acoustic signals with propagation distances of 300km and 400km

Subsequently, as shown in Figure 15, based on the measured sound velocity profile, this article uses the Kraken model to calculate the group velocities of the top ten normal modes within 100Hz, and obtains the group velocity dispersion curve. The average group velocity in the 30Hz to 80Hz frequency band is used as the equivalent average group velocity. Finally, as shown in Figure 18, based on the estimated arrival time difference of the first and second normal modes from the measured signals and the equivalent average group velocity obtained from model simulation, the estimated source distance of the 200km sound signal is 205.3km, the 300km sound signal is 308.9km, the 400km sound signal is 378.6km, and the 518km sound signal is 382.4km. The estimation results of the sound source distance for the first three signals are relatively accurate and have a small error, while the estimation results for the fourth signal have a large error. This is because when the propagation distance is far, there is a horizontal change in the dual-channel SSP, and the equipment group velocity calculated by the modes does not consider the horizontal change in the SSP.

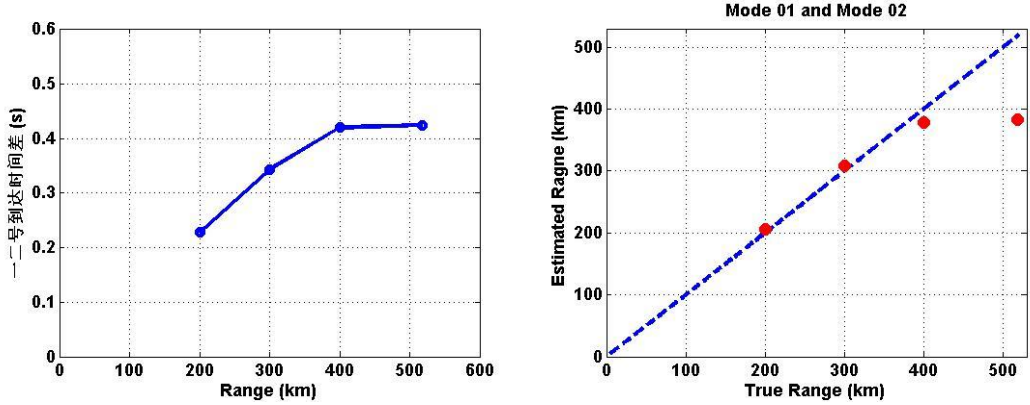

Figure 18 Estimation of the Time Difference of Arrival of Normal Modes and Corresponding Sound Source Distance Based on Measured Signal Estimation

## VII.    CONCLUSION

This article analyzes the sound propagation characteristics of the Chukchi Plateau in the Arctic based on the normal mode theory. Overall, the Chukchi Plateau has a dual-channel SSP consistent with the Canadian Basin area, and also has the characteristics of a typical Arctic half-waveguide SSP. In this kind of marine environment, low-frequency broadband acoustic signals exhibit a typical dispersion structure of Arctic sound propagation, which is a refracted normal mode dispersion structure with gradually increasing modal frequency over time, as well as a cross phenomenon of modal dispersion curves caused by dual-channel SSPs. At the same time, the dispersion structure of the acoustic signal is influenced by the seabed terrain, which blocks the forward propagation of the lower frequency parts of each mode in the shallower seabed terrain. In addition, the sound signals received at surface depths exhibit the properties of normal modes, while the sound signals at medium and large depths exhibit the form of multipath sound rays. The above research viewpoints have been validated using sound propagation simulation and experimental data. In addition, the article also utilized sound propagation to demonstrate that in marine environments with significant changes in seabed terrain, the dispersion structure obtained from the simulation of seabed terrain unchanged environment can be used as a substitute, and the two are basically consistent.

This article uses measured experimental data to demonstrate that under the condition of a dual-channel SSP in the Chukchi Plateau sea area, when the seawater depth is large, the received broadband sound signal had complete dispersion structure. By intercepting the middle part of the sound signal and using the warping transformation of refractive normal modes, the time-domain signal separation and dispersion structure extraction of normal modes can be achieved. When the seabed terrain is complex and the received broadband acoustic signal does not have a complete dispersion structure, modal separation can also be achieved by intercepting partial signals. The above research results indicate that in the future, extracting dispersion structures can be used to carry out the next step of marine environmental acoustic inversion and sound source localization research.

Based on the broadband acoustic signal characteristics in the dual-channel SSP environment of the Chukchi Plateau, this article proposes a sound source distance estimation method based on matching the duration of the dispersion structure of the normal mode, under the condition that the acoustic signal has a complete dispersion structure. This method combines the duration of the dispersion structure extracted from the time-domain waveform and the equivalent group velocity calculated based on the normal mode model to achieve sound source distance estimation, This method is less affected by changes in the surface sound velocity profile, and the effectiveness of the method has been well verified using measured experimental data in the article. Under the condition that the sound signal does not have a complete dispersion structure, this article proposes another sound source distance estimation method based on the dispersion structure characteristics of the dual-channel SSP, which is based on the modal arrival time difference obtained from time-domain waveforms. This paper also verifies the effectiveness of the method using measured experimental data, but it is susceptible to horizontal changes in the sound velocity profile of surface seawater.

Additionally, this article has some shortcomings. For example, the air gun sound signal received at the Chukchi Plateau this time did not have corresponding accurate sound source location and emission time information. This experiment also did not have accurate measured seabed topography and sound velocity profile information. The experiment did not conduct large-scale sound velocity profile measurement work on the Chukchi Plateau to obtain details of the changes in seawater sound velocity profile with longitude and latitude. Accurate sound velocity profile information is beneficial for improving the estimation results of the second sound source distance estimation method in the article. However, this article provides a good demonstration using a seismic exploration experiment at the Chukchi Plateau, which demonstrates how to extract underwater acoustic waveguide information from common air gun sound signals in the Arctic Ocean. In future work, a large amount of collected air gun sound signals can be used to study the Arctic marine environment and invert the Arctic marine environment, thus achieving the observation of the Arctic Ocean using air gun sound signals.

## ACKNOWLEDEMENTS

This research was supported in part by the National Key Research and Development Program of China under Grant, in part by the National Natural Science Foundation of China under Grant, in part by the Scientific Research Foundation of Third Institute of Oceanography, State Oceanic Administration under Grant, and in part by the Opening Foundation of State key laboratory of acoustics under Grant. Special thanks to all the staff involved in the deployment and recycling of the equipment during the Experiment.

# REFERENCES

1. Finn B. Jensen, William A. Kuperman, Michael B. Porter, Henrik Schmidt, Computational Ocean Acoustics, Second Edition, Springer, New York, 2011
2. Michael B. Porter, The KRAKEN Normal Mode Program (DRAFT)
3. Michael D. Collins, User's Guide for RAM Versions 1.0 and 1.0p
4. Kelly A. Keen, Bruce J. Thayre, John A. Hildebrand, Sean M. Wiggins, Seismic airgun sound propagation in Arctic Ocean waveguides, Deep-Sea Research Part I, 2018.09.003: 1-9
5. Yang, T. C., 1984, Dispersion and ranging of transient signals in the Arctic Ocean, J. Acoust. Soc. Am. 76, 262-273
6. T. C. Yang, A method of range and depth estimation by modal decomposition, The Journal of the Acoustical Society of America, 1987, 82, 1736-1745
7. Henry Kutschale, Long-range sound transmission in the arctic ocean, Journal of Geophysical Research, 1961, 66: 2189-2198
8. Marsh H. W., and Mellen R. H., Underwater sound propagation in the arctic ocean, J. Acoust. Soc. Am., 1963, 35: 552-563
9. Buck B. M. and Greene C. R., Arctic deep-water propagation measurements, J. Acoust. Soc. Am., 1964, 36: 1526-1533
10. Arthur B. Baggeroer, Jon M. Collis, Transmission loss for the Beaufort Lens and the critica frequency for mode propagation during ICEX-18, The Journal of the Acoustical Society of America, 2022, 151, 2760-2772
11. Timothy F. Duda, Weifeng Gordon Zhang, and Ying-Tsong Lin, Effects of Pacific Summer Water layer variations and ice cover on Beaufort Sea underwater sound ducting, The Journal of the Acoustical Society of America, 2021, 149, 2117-2136
12. Megan S. Ballard, Mohsen Badiey, Jason D. Sagers, et al, Temporal and spatial dependence of a yearlong record of sound propagation from the Canada Basin to the Chukchi Shelf, The Journal of the Acoustical Society of America, 2020, 148, 1663-1680
13. Peter F. Worcester, Mohsen Badiey and Hanne Sagen, Introduction to the special issue on ocean acoustics in the changing arctic, The Journal of the Acoustical Society of America, 151, 2787-2790
14. EeShan C. Bhatt, Oscan Viquez and Henrik Schmidt, Under-ice acoustic navigation using real-time model-aided range estimation, The Journal of the Acoustical Society of America, 2022, 151, 2656-2671
15. Gaute Hope, Hanne Sagen, Espen Storheim, Halvor Hobak, and Lee Freitag, Measured and modeled acoustic propagation underneath the rough Arctic sea-ice, The Journal of the Acoustical Society of America, 2017, 142, 1619-1633
16. Final Environmental Assessment/Analysis of Marine Geophysical Surveys by R/V Sikuliaq in the Arctic Ocean, Summer 2021
17. Ethan H Roth, John A. Hildebrand, and Sean M. Wiggins, Underwater ambient noise on the Chukchi Sea continental slope from 2006-2009, The Journal of the Acoustical Society of America, 2012, 131, 104-110
18. Qi Yu-Bo, Zhou Shi-Hong, Zhang Ren-He, Warping transform of the refractive

normal mode in a shallow water waveguide, Acta Physica Sinica, 2016, 65, 134301-1-9(in Chinese)
19. Aaron Thode, Katherine H. Kim, Charles R. Greene, and Ethan Roth, Long range transmission loss of broadband seismic pulses in the Arctic under ice-free conditions, The Journal of the Acoustical Society of America, 2010, 128, EL181-187
20. Kevin LePage and Henrik Schmidt, Modeling of low-frequency transmission loss in the central Arctic, The Journal of the Acoustical Society of America, 1994, 96, 1783-1795
21. Murat Kucukosmanoglu, John A. Colosi, Peter F. Worcester, Matthew A. Dzieciuch, and Daniel J. Torres, Observations of sound-speed fluctuations in the Beaufort Sea from summer 2016 to summer 2017, The Journal of the Acoustical Society of America, 2021, 149, 1536-1548
22. A. G. Litvak, Acoustics of the deepwater part of the Arctic Ocean and Russia's Arctic shelf, Herald Russ. Acad. Sci, 85, 2015, 239-250
23. M. A. Spell, R. S. Pickart, M. Li, M. Itoh, P. Lin, T. Kikuchi and Y. Qi, Transport of Pacific Water into the Canada basin and the formation of Chukchi slope current, J. Geophys. Res: Oceans, 123, 2018, 7453-7471
24. Peter N. Mikhalevsky, and Alexander N. Gavrilov, Acoustic thermometry in the Arctic Ocean, Polar Research, 2001, 20, 185-192
25. Alexander N. Gavrilov, Peter N. Mikhalevsky, low-frequency acoustic propagation loss in the Arctic Ocean: results of the Arctic Climate Observations using underwater sound experiment,
26. The Coordinated Arctic Acoustic Thermometry Experiment- CAATEX
27. Bonne Julien, Aaron Thode, Dana Wright and Ross Chapman, Nonlinear time-warping made simple: A step-by-step tutorial on underwater acoustic modal separation with a single hydrophone, 2020, 1897, 147
28. Julien Bonnel, Aaron M. Thode, Susanna B. Blackdwell and Kathering Kim, A. Michael Macrander, Range estimation of bowhead whale (Balaena mysticetus) calls in the Arctic using a single hydrophone, Journal of the Acoustical Society of America, 2014, 136, 145
29. Julien Bonnel, Stan dosso, N. Ross chapman, Bayesian geoacoustic inversion of single hydrophone light bulb data using warping dispersion analysis, Journal of the Acoustical Society of America, 2013, 134, 120
30. Niu haiqiang, Renhe Zhang, and Zhenglin Li, Theoretical analysis of warping operators for non-ideal shallow water waveguides, Journal of the Acoustical Society of America, 2014, 136, 53,
31. Deming Zhang, Zhengling Li, Renhe Zhang, Inversion for the bottom geoacoustic parameters based on adaptive time-frequency analysis, ACTA ACUSTICA, 2005, 30(5), 415-419(in Chinese)